\documentclass[usenatbib]{mn2e}
\usepackage{graphicx,epsfig,amsmath,soul,natbib}
\usepackage[normalem]{ulem}
\usepackage{times}
\usepackage{abbrev}
\newcommand{\ltaraw}{$\; \buildrel < \over \sim \;$}
\newcommand{\lta}{\lower.5ex\hbox{\ltaraw}}
\newcommand{\gtaraw}{$\; \buildrel > \over \sim \;$}
\newcommand{\gta}{\lower.5ex\hbox{\gtaraw}}
\renewcommand{\v}[1]{\ensuremath{\mathbf{#1}}}

\title[The Sunyaev-Zeldovich background]{The Sunyaev-Zeldovich background}

\author[G.P.~Holder et al.]
{Gilbert P. Holder$^1$\thanks{CIAR Scholar~and~Canada Research Chair in Cosmological Astrophysics}, 
    Ian G. McCarthy$^{2}$, A. Babul$^3$\thanks{E-mail: holder@physics.mcgill.ca (GPH), 
    i.g.mccarthy@durham.ac.uk (IMG), babul@uvic.ca (AB)}\\
  $^1$Department of Physics, McGill University, Montreal, QC H3A 2T8, Canada \\
  $^2$Department of Physics, University of Durham, South Road, Durham, DH1 3LE, UK\\
  $^3$Dept. of Physics \& Astronomy, University of Victoria,
    Elliott Building, 3800 Finnerty Rd., Victoria, BC V8P 1A1, Canada}

\date{draft version \today}

\pagerange{\pageref{firstpage}--\pageref{lastpage}} \pubyear{2005}

\def\msun{{\rm\; M}_\odot}

\begin{document}

\label{firstpage}

\maketitle

\begin{abstract}
The cosmic background due to the Sunyaev-Zeldovich (SZ) effect is
expected to be the largest signal at mm and cm wavelengths
at a resolution of a few arcminutes. We investigate some simple
statistics of SZ maps and their scaling with the normalization
of the matter power spectrum, $\sigma_8$, as well as the effects
of the unknown physics of the intracluster medium on these statistics.
We show that the SZ background
provides a significant background for SZ cluster searches, with the
onset of confusion occurring around $10^{14} h^{-1} M_\odot$ in
a cosmology-dependent way, where confusion is defined as
typical errors in recovered flux larger than 20\%. 
The confusion limit, corresponds
to the mass at which there are roughly ten clusters per square degree, with
this number nearly independent of cosmology and cluster gas
physics.  Typical errors grow quickly as lower
mass objects are included in the catalog.

We also point out that there is nothing in particular about
the rms of the filtered map that makes it especially well-suited for capturing
aspects of the SZ effect, and other indicators of the
one-point SZ probability distribution function are at least as well
suited for the task. For example, the full width at half maximum of the 
one point probability distribution has a field-to-field scatter that
is about $60\%$ that of the $rms$. 

The simplest statistics of SZ maps are largely unaffected by cluster 
physics such as
preheating, although the impact of preheating is clear by eye in the maps.
Studies aimed at learning about the physics of the intracluster medium will
apparently require more specialized statistical indicators.

\end{abstract}

galaxies: clusters: general ; galaxies: intergalactic medium ; cosmology: cosmic microwave background
\section{Introduction}

The properties of the cosmic microwave background (CMB), and particularly its anisotropies,
are a treasure trove of information about the fundamental cosmological parameters that define
the large-scale structure of the Universe and about the early Universe physics that set the
stage for the emergence of the present-day cosmic structures.  As CMB experiments
reach higher sensitivity (approaching $1\;\mu{\rm K}$) and higher resolution
(approaching 1 arcminute) across a wide range of frequencies (from 20 to 900 GHz), they
also hold forth the promise of providing new windows for probing the  distribution of
matter at redshifts $z\ll 1000$.  This material is the dominant source of the temperature
fluctuations in the CMB on scales less than 4 arcminutes.

The largest source of these small-scale anisotropies at the mm and cm wavelengths is expected
to be the Sunyaev-Zeldovich (SZ) distortion from clusters of galaxies \citep{sunyaev72,
birkinshaw99,carlstrom02}.  This distortion is caused by the inverse-Compton
scattering of CMB photons by hot intracluster gas, and it has a
characteristic spatial and spectral signature in CMB sky maps.  
The magnitude of
the SZ effect from a galaxy cluster is determined by the integrated gas pressure along the 
line of sight;  SZ anisotropies thus have the potential to yield valuable insights 
regarding the physical processes at play within clusters, especially those that have shaped 
the spatial and thermodynamic properties of the diffuse baryons. 

The current measurements of the small-scale anisotropy  by instruments such
as the Cosmic Background Imager (CBI) \citep{mason03} and the Berkeley-Illinois-Maryland
Array (BIMA) \citep{dawson02} are broadly consistent \citep{komatsu02,holder02a,bond05}
with levels expected from the SZ effect, though the slight discrepancy in the normalization
of the mass fluctuations $\sigma_8$ implied by these measurements versus those implied by
other cosmological probes may be a harbinger of interesting times ahead \citep{dore04}.
Current experiments such as the Sunyaev-Zeldovich Array (SZA) \citep{muchovej06}, 
the Atacama Cosmology Telescope (ACT) \citep{kosowsky06}, 
and the South Pole Telescope (SPT) \citep{ruhl04} 
ought to be able to able to image the
small-scale anisotropies with exquisite signal-to-noise.

Given their importance, considerable attention has been devoted to identifying approaches for 
quantifying, analyzing and interpreting these small-scale anisotropies in the CMB maps.  Typically 
these have involved either constructing source counts as a function of source size or source flux 
density (e.g.~\citealp{barbosa96}) or computing the SZ power spectrum (e.g.~\citealp{cole88}).
(We refer the reader to the review by \citealp{carlstrom02} for a more detailed discussion
of these two approaches as well as an extensive list of related references.)
A hybrid approach was outlined in \citet{diego04}.

In this paper, we examine both these approaches.  We show that the SZ background is 
effectively an irreducible contaminant that must be factored into the scan strategies for upcoming
SZ survey experiments.   We determine the mass scale below which SZ counts will be subject
to confusion and show that the current generation of SZ experiments will be very close to being
confusion limited if $\sigma_8=0.9$.  We also investigate the ``bandpower'' approach
to characterizing the SZ power spectrum.   At first glance, this approach may appear to be a
rather blunt measure, raising questions about its efficacy for providing useful information about the 
underlying sources of the anisotropies.

For these investigations, we generate synthetic maps of the SZ effect 
as we vary cluster gas physics and a single cosmological parameter,
the normalization of the matter density power spectrum ($\sigma_8$).  Our map-making
algorithms and underlying assumptions are discussed in \S~\ref{mapmaking}.   Our methodology is
specifically tailored to capture the effects of projection and clustering of galaxy
clusters but not substructure and asphericity within the clusters.  Admittedly, the cluster profiles
are not very realistic but since we are primarily concerned with the
SZ background, we do not believe that this limitation is critical.  On the contrary,
since our maps are free of the ambiguities associated with determining whether one is
looking at several objects in projection or a single object with multiple components,
we are able to identify cleanly the influence of projection and clustering on the SZ maps.
Furthermore, the regularity of cluster populations in X-ray properties \citep{nagai07}, 
which should be
much more sensitive to substructure, suggests that substructure and asphericity are not likely
to be large problems.

We use our SZ maps to investigate the SZ background as a source of contaminant
for source counts extracted from a wide-field survey.  Precisely determined SZ
source counts are potentially  sensitive tools for studying of the growth of
structure over cosmic time, for determining the fundamental cosmological parameters,
and for constraining unknown quantities such as the dark energy equation of state.
The cosmological power of these surveys is crucially dependent \citep{majumdar03} on the 
relationship between the observable quantities (such as, flux and angular size) and
the relevant intrinsic properties of the source population (such as mass). A 
significant strength
of SZ source counts as a cosmological tool, the lack of redshift-dependent dimming, is also
a weakness: 
the SZ sky will also consist of superposition of numerous weak distortions associated
with largely unresolved clusters spanning a range of redshifts and mass scales.
These background fluctuations can significantly distort the observable quantities
associated with identified sources.  In \S~\ref{confusion}, we comment on the resulting
consequences for future SZ wide-field cluster surveys.

Next we look at current approaches to characterizing the SZ power spectrum.
In reporting a power spectrum on a given scale,
a complex map of numerous sources of varying amplitudes and scales is reduced to
a single number, a ``band-power'', characterizing the rms fluctuation amplitude in a
SZ map filtered through some bandpass window centered about the scale of interest.
For the primary CMB anisotropies, this type of characterization is well motivated,
since the fluctuations are believed to Gaussian random and the statistical nature of
a Gaussian random field is completely specified by its two-point correlation function
or equivalently, its power spectrum.  SZ fluctuations, however, are not Gaussian
random and characterizing them via a power spectrum only makes sense
in the limit of a large number of sources leading to the bulk of the anisotropy,
where one might hope that the map becomes nearly Gaussian through the central
limit theorem.  Indeed, as we demonstrate in \S~\ref{mapstats}, 
there is nothing about the rms or the variance of a filtered map that identifies it as
especially well-suited for characterizing the SZ anisotropies. The variance has the nice
property that it adds with the noise variance in a way that is independent of the
shape of the probability distribution, but as the new generation of experiments begins
to image the SZ background at high signal to noise this is not necessarily a large
advantage. In fact, there are several alternate quantities derived from the one-point 
SZ fluctuation probability distribution function that allow for somewhat 
better cosmological discrimination, even in the simplest case where we restrict ourselves
to reducing the richness of the SZ probability distribution 
\citep{zhang07}
to a single number.

By restricting our study to one-point statistics of idealized cluster sources we have
eliminated many real-world complications that are both important and confusing, such
as substructure, detector noise, contamination by primary and secondary CMB anisotropies and 
foregrounds. However, this simplicity allows a clearer understanding of the impact
of effects such as projection effects that would be difficult to clearly identify and isolate
in full numerical simulations. Increasing the realism of the sky maps will be a necessary
next step to be able to connect to real experiments in a meaningful way. However, we expect
the results below to be robust.

\begin{figure}
\begin{center}
\includegraphics[height=9cm]{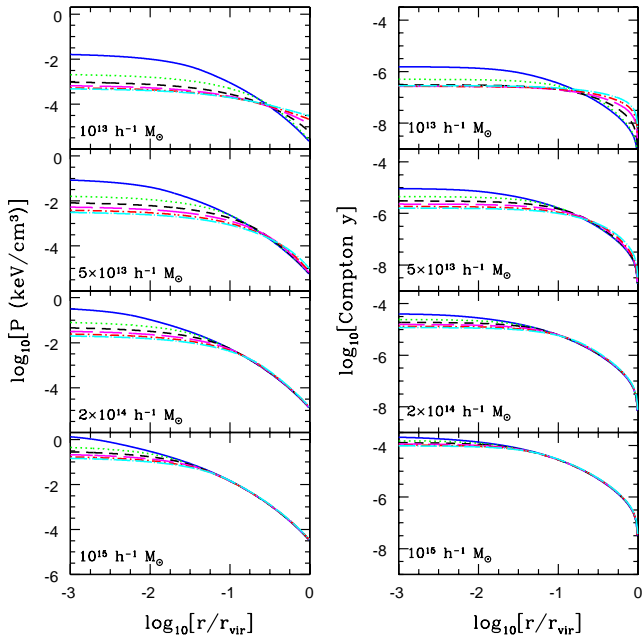}
\caption{
The left column shows, from top to bottom, the spherical pressure profiles for clusters of masses 
$1\times 10^{13}\;h^{-1}\msun$, $5\times 10^{13}\;h^{-1}\msun$, $2\times 10^{14}\;h^{-1}\msun$, 
and $1\times 10^{15}\;h^{-1}\msun$.  In each panel, we show pressure profiles corresponding to 
$S_\circ =10$, $100$, $200$, $300$, $400$ and $500$ keV cm$^2$ (in descending order at $r/R_v =0.01$).
The right column shows the corresponding projected Compton $y$ profiles.  For a given cluster,
increasing $S_\circ$ has two effects on the cluster's Compton $y$ profile.  The amplitude of the $y$ 
profile in the inner regions of the cluster diminishes while the signal in the outer regions
increases.  Both effects are the most pronounced for low cluster masses and high $S_\circ$. The lower
mass clusters have corresponding shallower potential wells and the gas in these systems is more
susceptible to rearrangement in response to feedback.
}
\label{fig:press-y}
\end{center}
\end{figure}

\section{Mapmaking Methodology}\label{mapmaking}

We opt for an approach that uses semi-analytic methods to generate halo catalogs 
as a function of cosmological parameters, and use a well-studied analytic model for the 
distribution of diffuse gas in groups and clusters to populate the halos with baryons 
in a realistic fashion.
Mechanically, this procedure is similar to that done by \citet{kay01}, who used the Hubble
Volume simulations to generate halo catalogs and will miss the effects of diffuse gas 
that are captured using hydrodynamical simulations \citep{white02,dasilva00}. 
Most of the anisotropy comes from bound objects \citep{dasilva01}, although there are
significant contributions to both the mean Compton $y$ and to the kinetic SZ effect
expected from unbound gas .

We use the Pinocchio formalism \citep{monaco02} to generate the halo catalogs.  Starting with 
a realization of the initial linear density field within some specified volume, this algorithm
not only predicts the mass function and the merger histories of the collapsed halos within the 
volume, but also provides information about their spatial locations.  The results have been 
shown to be in excellent agreement with the results of full $N$-body simulations. Pinocchio, however,
is computationally several orders of magnitude faster.

For the present study, we restrict our consideration to a reference cosmological model 
characterized by $\Omega_m=0.3$, $\Omega_\Lambda=0.7$ and $h=0.7$ but allow $\sigma_8$
to span the range from 0.6 to 1.0. 
For each value of $\sigma_8$, we generate six different simulation cubes of 256 $h^{-1}$ 
comoving Mpc on side, and output the corresponding halo catalogs at redshift intervals 
corresponding to 256 $h^{-1}$ Mpc in comoving distance.  These simulation volumes are 
stacked to generate an integrated catalog of halos along a line-of-sight that correctly
accounts for the significant evolution of cosmic structure as function of look-back
time.  Rather than simply stacking simulations corresponding to 
one realization, we randomly choose between one of the six volumes at each 
redshift, randomly select between one of three possible orientations and further apply a
random periodic global translation to its content.  Once the stack is assembled, we compile a list
of all halos that fall within a $2^{\circ} \times 2^{\circ}$ field as seen by an observer 
at $z=0$.   For each value of $\sigma_8$, we construct 100 realizations of the
the $2^{\circ} \times 2^{\circ}$ sky field, and we only include 
halos with masses above $10^{13} h^{-1} M_\odot$ in the catalog. The mass threshold does
not have a significant impact on the results. The SZ power spectrum is dominated by
galaxy clusters of masses on the order of $10^{14} h^{-1} M_\odot$
\citep{holder02b,atrio99}. This field size is comparable in width to the ``ACT Strip'' (although
much shorter in length), and of the same order as a typical SZA field. 

Next, we populate each halo in the final catalog with  hot diffuse baryons according to the 
analytic model of \citet{mccarthy03a}.  These models provide a simple, physically intuitive 
description of the intracluster medium subject to both heating due to accretion and infall, as
well as non-gravitational heating events phenomenologically associated with, for example, 
starburst and AGN activity at the time when cluster galaxies are form, or AGN outbursts 
triggered by inflow of cooling gas once the cluster approaches a relaxed configuration.

In the standard hierarchical framework, the thermodynamic
properties of the intracluster medium (ICM) are determined purely by heating due to accretion 
shocks and compression, and by radiative cooling.  It has been known for a decade, however,
that cluster models that include only these processes  fail to reproduce both the observed 
stellar-to-gas mass ratio and the mean observed X-ray properties of the clusters 
(e.g.~\citealp{lewis00, balogh01, babul02, voit02, lin03, mccarthy04}). For instance,
the observed X-ray luminosity-temperature ($L_x-T_x$) relation is steeper than predicted.
Also, recent analytic and numerical investigations show that neither the cosmic halo-to-halo
variations in the detailed structure of dark matter distribution of the cluster halos  
nor the effects of mergers can account for the relatively  large scatter  in the 
$L_x-T_x$ relation \citep{rowley04, mccarthy04, balogh06, poole07}.  On the other hand, models that, in 
addition to radiative cooling and accretion-related heating, also include early heating 
associated with AGN activity can account for most of the observed X-ray/SZE/optical
scaling relations (c.f. \citealp{babul02, nath02, mccarthy02, mccarthy03b, mccarthy04, 
ostriker05, roychowdhury05, poole07, mccarthy07}) but also
for the {\em real} scatter in $L_x-T_x$ \citep{mccarthy04, balogh06}.   Additionally, this class
of models can also resolve one of the most persistent problems in galaxy formation:
the production of far too many high luminosity galaxies than observed \citep{benson03}.  AGN feedback 
naturally produces an anti-hierarchical quenching of star formation
in large galaxies, consistent with observations of galaxy down-sizing, due to the relatively
late peak in ($z\approx 2$) in nuclear activity \citep{scannapieco05}.

Purely gravitational processes produces entropy profiles of the form $S(r)\propto r^{1.1}$ outside 
the cluster cores. This behaviour has been noted in numerical simulation studies of clusters 
\citep{lewis00, kay04, voit05} and has also been observed in a wide range of clusters \citep{pratt05, 
piffaretti05, donahue05, mccarthy07}.  In the models of \citet{mccarthy03a}, the extent of AGN heating is parameterized 
through the modification of the gas entropy profile, specifically through the introduction of an 
entropy core of amplitude $S_\circ$.  The gas distribution in a halo is specified by the specific
value of $S_\circ$ and requiring that the gas is in hydrostatic equilibrium.  Under the assumption
that $S_\circ$ varies from cluster to cluster, one can interpret the results as a providing a snapshot 
in time of the gas distribution in a halo.  Unlike the more sophisiticated models of 
\citet{mccarthy04, mccarthy07}, those described in \citet{mccarthy03a} do not track the evolution of 
$S(r)$ as a function of time.  We have opted to used the more simpler models because by 
allowing for a variation in $S_\circ$, we are able to span the full range of plausible entropy 
profiles and gas distributions without having to concern ourselves with details of the 
heating mechanism and of the timing of the heating events in individual clusters.  

\begin{figure}
\begin{center}
\includegraphics[width=84mm]{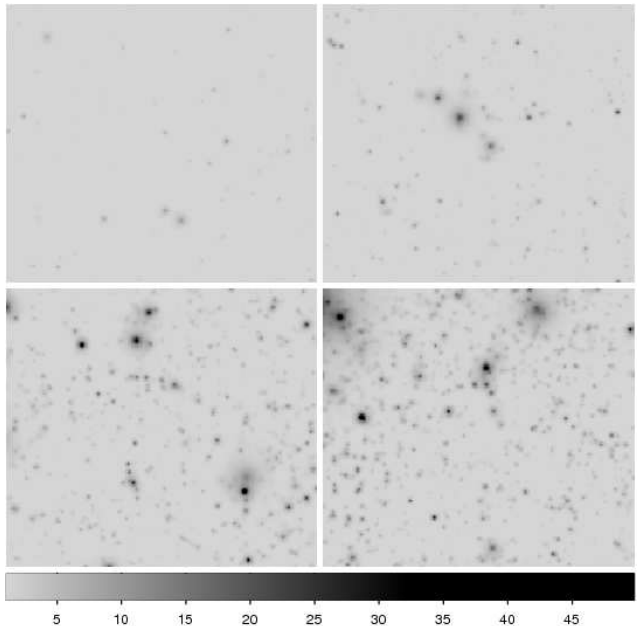}
\caption{
Examples of maps for different choices of $\sigma_8$. Each map is a randomly chosen
realization and shows a square that is 0.85 degrees on a side. The values of $\sigma_8$
are 0.6 (top left), 0.7 (top right), 0.8 (bottom left), and 0.9 (bottom right) and the
cluster model is chosen to have $S_\circ = 200$ keV cm$^{2}$. The map resolution is
14" and the shown region is only the central part of a map that is 2 degrees on a side.
The units of the map are $10^6 \times y$, {\em not} temperature units. To convert to 
Rayleigh-Jeans temperature in units of $\mu K$, multiply by 5.46.
}
\label{fig:skymap}
\end{center}
\end{figure}

The nomalization, slope (shape)  and the scatter in the various SZ effect and X-ray scaling relations 
as well as the variations in the shape of X-ray surface brightness profiles of massive clusters 
suggest that $S_\circ$  spans the range from $\sim 0-10$ keV cm$^2$ to $\sim 400-500$ keV cm$^2$,
with the entropy profiles for the  $S_\circ = 10$ keV cm$^2$ models being in excellent agreement 
with the observed entropy profiles of the classical cool core clusters \citep{donahue05, mccarthy07}.
We account for this range by generating a series of spherical pressure profiles corresponding to 
different values of $S_\circ$ for each of the halos in our final catalogs and use these these to obtain
circularly symmetric Compon $y$ profiles characterized by cluster mass, redshift and entropy 
parameter $S_\circ$.  All of the physics of the SZ effect is contained in the Compton $y$ profiles: 
\begin{equation}
\label{eqn:compton-y}
y(\v{\theta})={\sigma_T \over m_ec^2} \int P_e(\v{r})dl,
\end{equation}
where \v{\theta} is the projected position from the cluster center, $\sigma_T$ is the 
Thompson scattering
cross section, $P_e(\v{r})\equiv n_e(\v{r})kT_e(\v{r})$ is the electron pressure of the ICM
at the three dimensional position \v{r} from the cluster center, and the integral is performed
over the line of sight ($l$) through the cluster.   For example, the change in the 
temperature of the CMB due to the SZ effect is \citep{carlstrom02}
\begin{equation}
\label{eqn:compton-deltaT}
\Delta T(\v{\theta})=T_{CMB}y(\v{\theta})\left[x{e^x +1 \over e^x -1} -4\right],
\end{equation}
where 
$x=h\nu/kT_{CMB}\approx \nu/56.85$ GHz, and $T_{CMB}=2.728$ \citep{fixsen96}.  In
the long wavelength regime (the Rayleigh-Jeans limit), $\Delta T(\v{\theta})\approx 
-2T_{CMB}y(\v{\theta})$

In Fig. \ref{fig:press-y}, we show both the spherical pressure profiles and
the corresponding projected Compon $y$ profiles for clusters of different masses and a range of 
$S_\circ$.  Typically, as $S_\circ$ is increased, the reorganization of the gas within the halo
leads to the reduction in the ampltitude and the flattening of Compon $y$ profiles in the 
cluster centers.  This reorganization also results in an increase in the amplitude of the
Compon $y$ profiles outside the cluster cores, but the fractional change is small and the effect
is negligible for clusters with $M \gta 10^{14}\; \msun$.  For smaller mass systems, however,
the rearrangement of the gas leads to a sizeable increase in Compon $y$ profiles over the bulk
of the cluster.

To construct our sky maps, we start with a given catalog of halo corresponding to a specific value of 
$\sigma_8$.  We use the positions of the halos in the catalogs to locate the cluster centers
in a $2^{\circ}\times 2^{\circ}$ sky field.  Next, we choose a value for $S_\circ$, select the associated
2D Compton $y$ profile for each of the halos in the catalog, and place these at corresponding
positions in the sky field to generate an SZ map.  We do this for all
catalogs, ending up with 100 realizations of the SZ sky maps for each ($\sigma_8$, $S_\circ$)
pair.  As already noted, we explore  $\sigma_8$ values ranging from $0.6$ to $1.0$, and for
each of $\sigma_8$, consider several different values of $S_\circ$.  We adopt the $\sigma_8=0.9$,
$S_\circ=200$ keV cm$^2$ model as our reference case.  Not all the values
of $\sigma_8$ in the range being considered are in concordance with either the constraints 
from the cluster abundance studies \citep{henry04} or the recent microwave background results
\citep{spergel03,spergel06}; nevertheless it is still instructive to consider the full range
in order to probe the dependence on $\sigma_8$.

Fig. \ref{fig:skymap} shows four SZ sky maps generated using a model where 
$S_\circ=200$ keV cm$^2$ and four different values of $\sigma_8$. The power of 
SZ statistics as a probe of $\sigma_8$ is apparent by eye from this figure, as
the amount of structure is a strong function of $\sigma_8$. These maps were selected
as the first realization of each ensemble and the central region was selected for
display.

\begin{figure}
\begin{center}
\includegraphics[width=84mm]{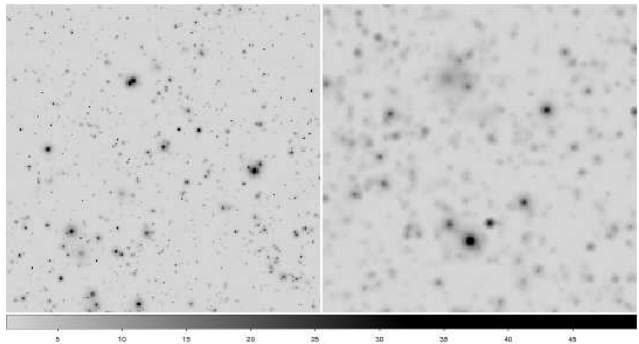}
\caption{
Examples of maps for extreme choices of entropy floor: 10 keV cm$^{2}$ (left)
and 500 keV cm$^{2}$ (right). Each map is a randomly chosen
realization and shows a square that is 0.85 degrees on a side. Both maps assume 
$\sigma_8=0.9$.
The map resolution is
14" and the shown region is only the central part of a map that is 2 degrees on a side.
The units of the map are $10^6 \times y$, {\em not} temperature units. To convert to 
Rayleigh-Jeans temperature in units of $\mu K$, multiply by 5.46.
}
\label{fig:entmap}
\end{center}
\end{figure}

There are two noteworthy items regarding
our SZ maps that we would like to draw attention to.  First, our sky maps are constructed
assuming that all the clusters have the same value of $S_\circ$.  In principle, we ought to have
allowed for a cluster-to-cluster variation in $S_\circ$ as suggested by the scatter in the 
SZ effect and X-ray observations; however, the shape of the distribution in $S_\circ$, 
much less its dependence on cluster mass and redshift, is not known.  To this end, we
have adopted the simplest possible distribution for $S_\circ$ and investigate the possible
implications for variations in $S_\circ$ by constructing maps with different values of $S_\circ$.
Second, although we have peculiar velocity information for each of the halos, we do not make use
of these in the present study.  Therefore, our maps only incorporate the thermal SZ effects,
not the kinetic SZ effects.  The latter are expected to approximately an order of magnitude
smaller and their inclusion would only have a marginal effect on the issues of interest
to us in the present study.

\begin{figure}
\begin{center}
\centerline{
\includegraphics[height=4.5cm]{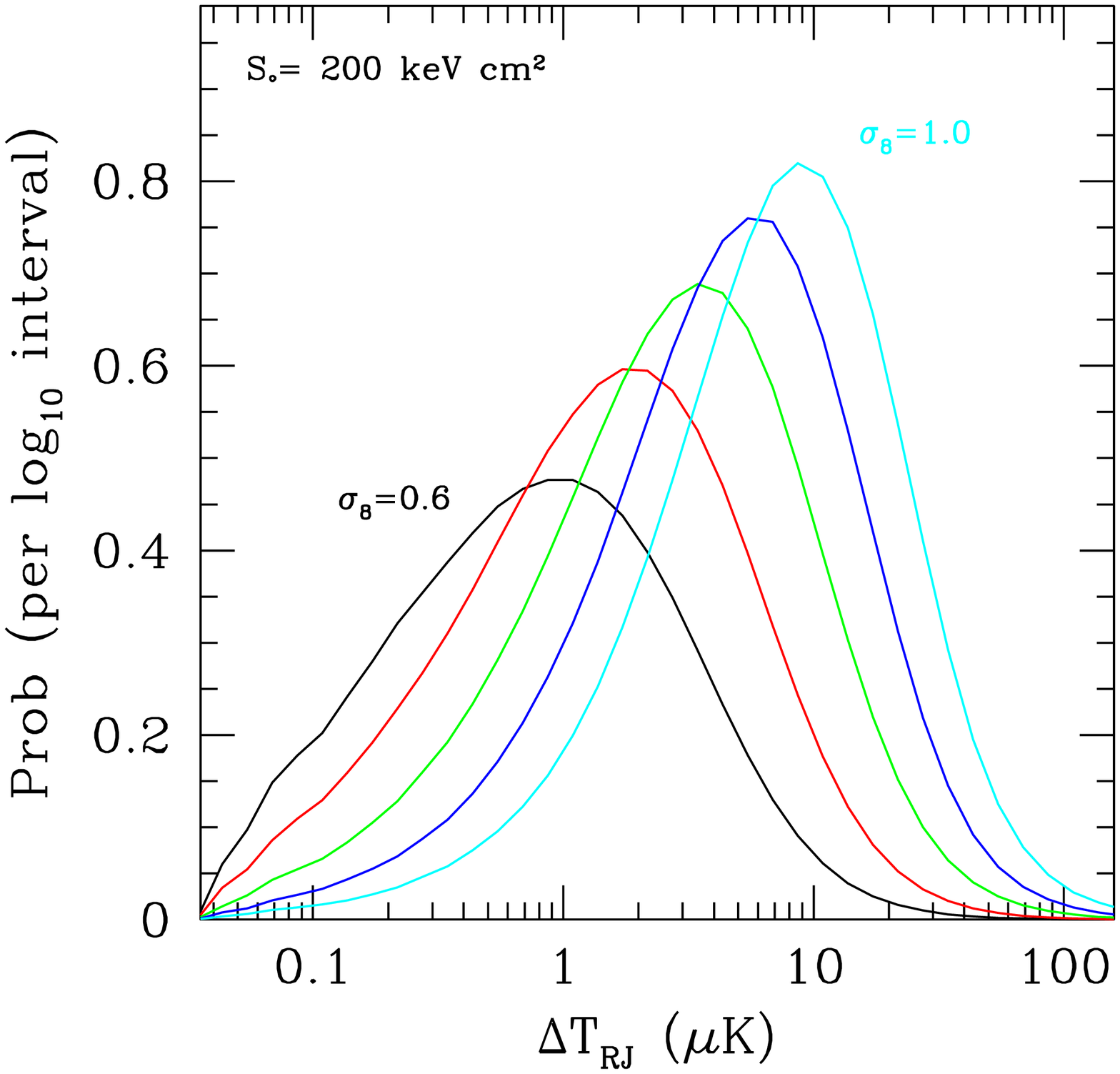} 
\includegraphics[height=4.5cm]{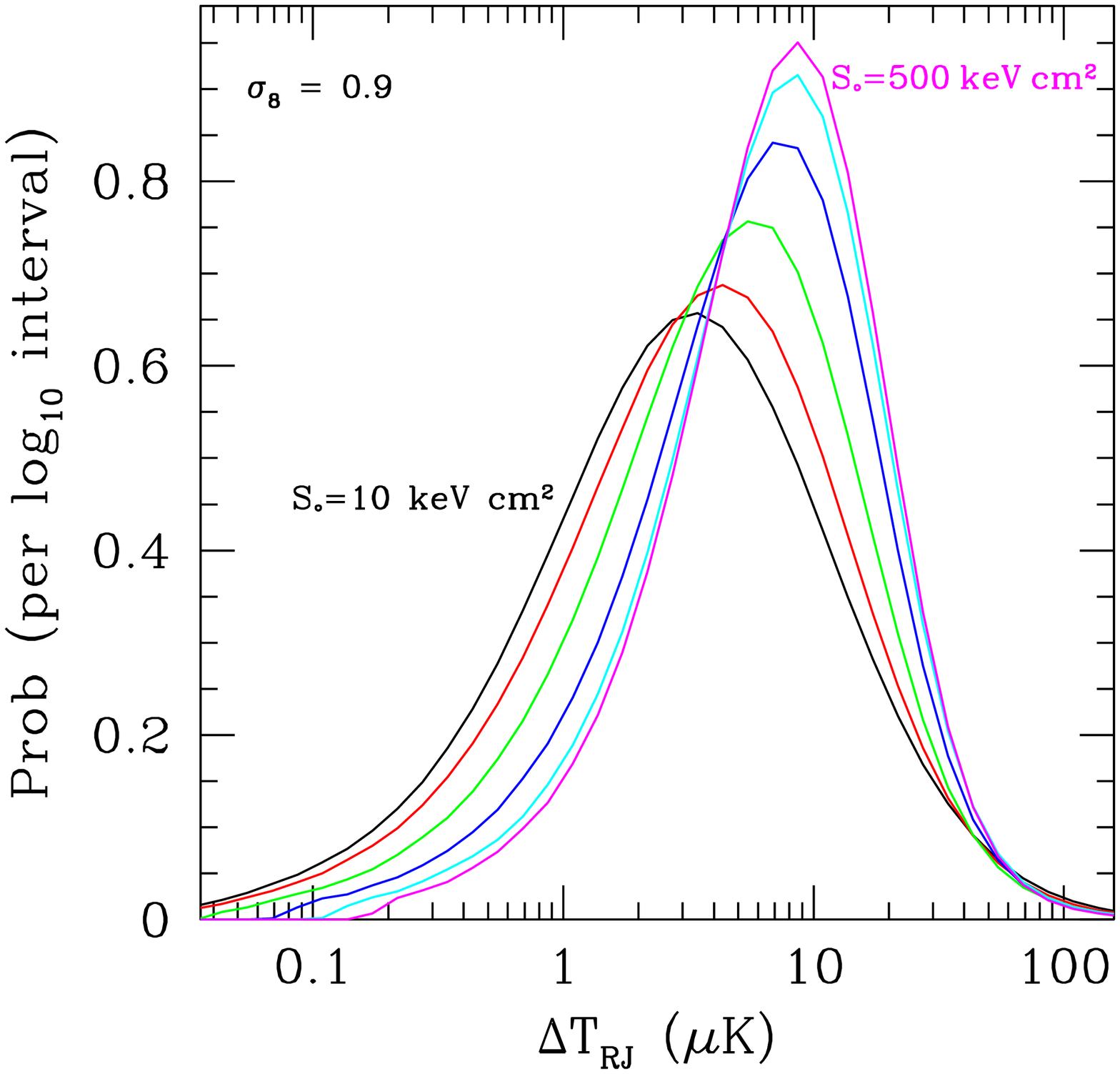}  }
\caption{
Distribution of pixel values for unfiltered maps (14" resolution) as a function of $\sigma_8$ (left)
and amount of preheating (right). There are some empty pixels (not shown) due to the 
neglect of unbound gas, with the fraction of empty pixels increasing with lower $\sigma_8$.
In the left panel, from left to right the curves show increasing $\sigma_8$ from 0.6 to
1.0 in steps of 0.1; the right panel shows increasing levels of preheating at a fixed value
of $\sigma_8=0.9$, from left to right showing $S_\circ=10,100,200,300,400,500$ keV cm$^{2}$.
}
\label{fig:unfiltered}
\end{center}
\end{figure}

The distribution of pixel values in the unfiltered maps captures 
some of the details apparent in the maps. As $\sigma_8$ increases 
the number of sources increases, as does the number of bright 
sources (see Figs. \ref{fig:unfiltered} and \ref{fig:entmap}).  
Increasing the value of $S_\circ$ has the effect of making 
clusters less centrally 
concentrated, leading to fewer very bright pixels but more pixels 
at intermediate values, as the outer regions of the clusters are 
somewhat enhance (as indicated by the crossover around 70 $\mu K$).  
The utility of SZ surveys as a cosmological 
probe is apparent in that the number of bright pixels is clearly 
very sensitive to $\sigma_8$ but not very sensitive to 
non-gravitational gas physics.  But it is also clear that 
high resolution $\mu K$ imaging of the SZ effect will provide 
insight into the detailed physical processes at play in clusters.

\begin{figure}
\begin{center}
\centerline{
\includegraphics[height=4.8cm]{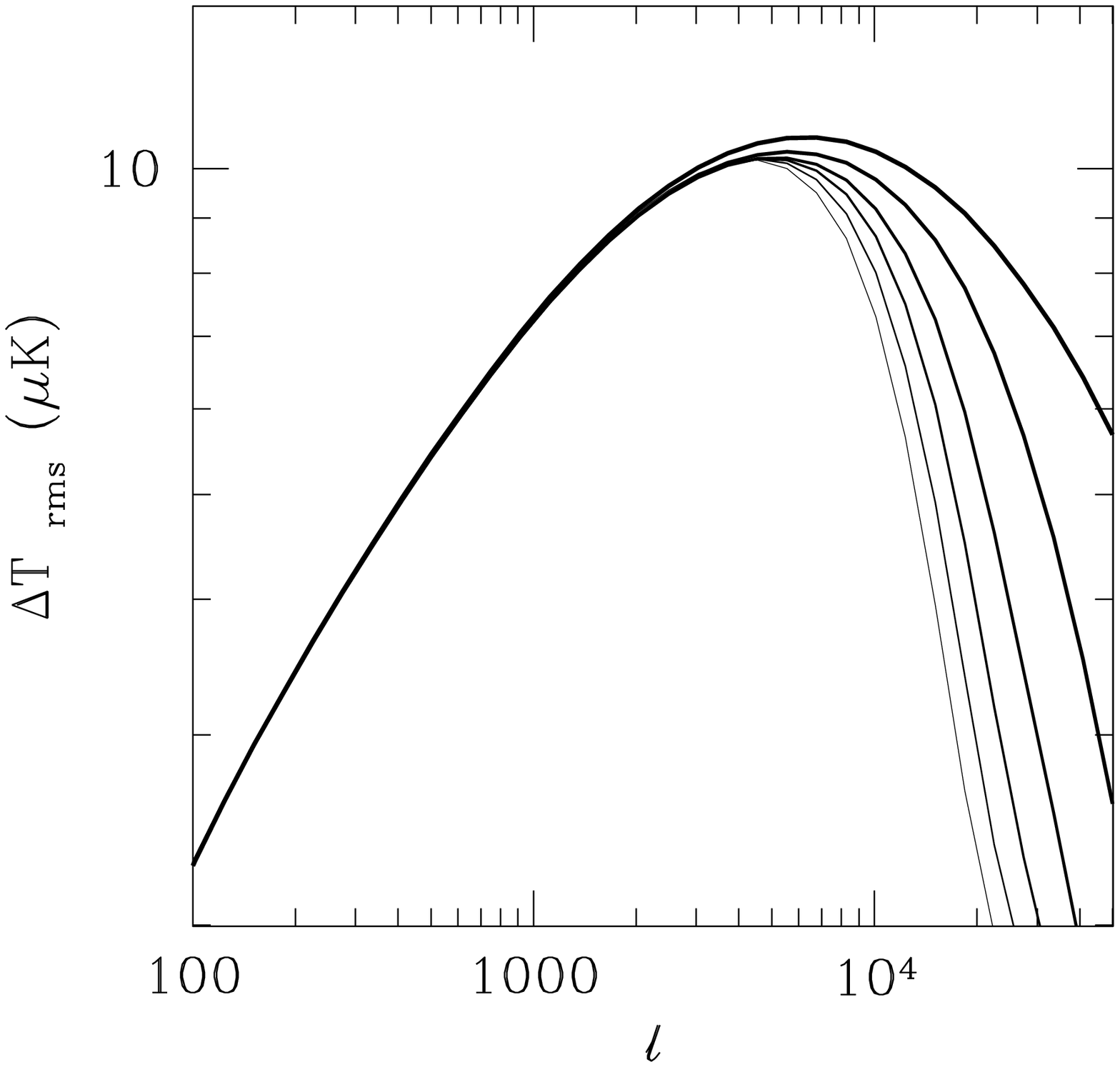}
\includegraphics[height=4.8cm]{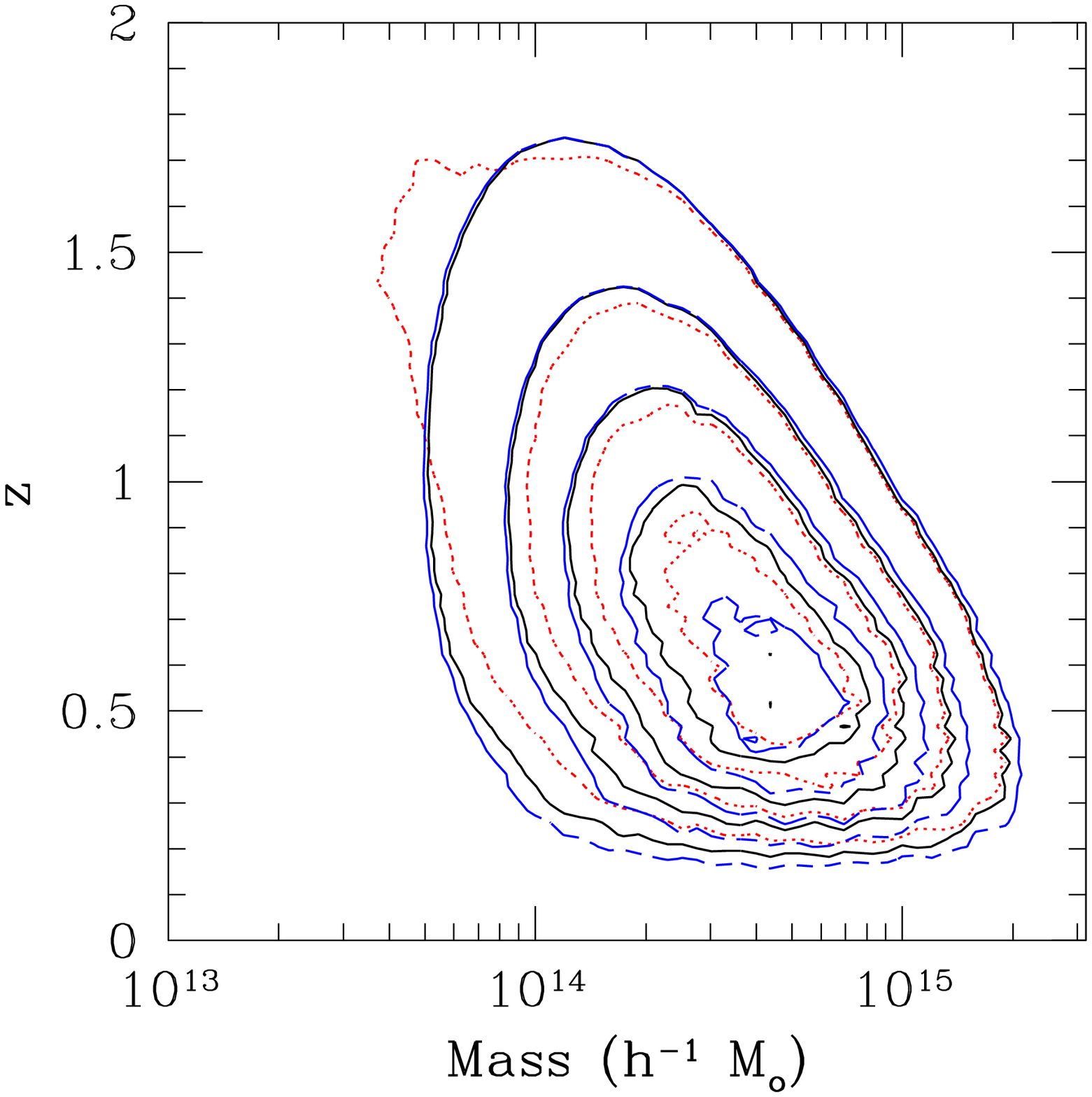}  }
\caption{The SZ power spectrum (left) and the differential contribution to 
SZ power spectrum, $d c_\ell/d \ln{M} dz$, at $\ell \sim 4500$ (right), for 
our models with $\sigma_8=0.9$ and varying levels of preheating. At high
$\ell$ the models from top to bottom have entropy floors of
10,100,200,300,400,500 keV cm$^{2}$ in the left panel;  the right panel
shows entropy floors of $S_\circ = 10$ (blue, dashed curve),  
200 (black, solid) and 400 keV cm$^2$ 
(red, dotted curve). See text for details on cluster models.  The contours in
the right panel are linearly spaced.
}
\label{fig:c_ell}
\end{center}
\end{figure}

\section{The SZ Angular Power Spectrum}\label{powerspectrum}

The distribution of the CMB temperature fluctuations  due to the SZ effect on
the celestial sphere is conventionally written as an expansion in terms of
the spherical harmonic functions,
\begin{equation}
\label{eqn:spherical}
{\Delta T\over T_{CMB}}(\theta, \phi)= \sum_{\ell, m} a_{\ell m}Y_{\ell m}(\theta,\phi)
\end{equation}
and quantified by the coefficients $a_{\ell m}$, or by the angular power 
spectrum of the SZ effect:
$C_\ell =\langle |a_{\ell m}|^2\rangle$. For small angles ($\ell \ga 50$) the flat sky
approximation should be sufficient, where the multipole expansion can be replaced
by a Fourier expansion \citep{white99} with $e^{2 \pi i \vec{u} \cdot \vec{x}}$
such that $2 \pi |u|=\ell$.

To gain some intuition for the SZ background,
we can use a Press-Schechter \citep{press74} approach to estimate the 
form of the SZ power spectrum.
Ignoring correlations between clusters, the power spectrum can be thought of as a superposition
of random sources. Each source has a profile in Fourier space that is modulated by a phase
factor due to its random position on the sky. At a given position in Fourier space, the 
superposition of many sources with random phases leads to a random walk in the real and
imaginary components. The power spectrum is thus given by
\begin{equation}
C_\ell = \int dz {dV \over dz} \int dM {dn \over dM} |\Delta \tilde{T}(\ell,M,z)|^2
\end{equation}
where $\Delta \tilde{T}$ is the Fourier transform of the SZ profile, $dV/dz$ is the
comoving volume per unit redshift and $dn/dM$ is the comoving number density of objects of
mass $M$. We use the number density indicated by fits to numerical simulations
\citep{jenkins01}. We obtain the Fourier transform of the SZ profile using the
$y$ profiles of the preheated models and a Hankel transform routine
to take advantage of the circular symmetry of the model clusters.

In left panel of Fig. \ref{fig:c_ell}, we plot angular SZ power spectra as rms
contribution per unit interval $[\ell (\ell +1)C_\ell /2\pi]^{1/2}$ for our reference
model ($\sigma_8=0.9$, $S_\circ=200$ keV cm$^2$) as well as the full range of
preheated models
with the same $\sigma_8=0.9$, with the $S_\circ$ parameter ranging from 
$10$ to 500 keV cm$^2$.  In the right panel, we show the differential 
contribution to the SZ power spectrum, $d c_\ell/d \ln{M} dz$, at $\ell \sim 4500$
for the same three models. The effects of feedback at moderate levels results
in the well-known damping of the SZ power on all scales but most pronouncedly,
on small scales \citep{holder01, lin04}.  
As illustrated 
in the right panel of Fig. \ref{fig:c_ell}, the power on intermediate scales
primarily comes from clusters of a few times $10^{14} h^{-1} M_\odot$ at 
moderate redshift and as a result these scales are little affected
by preheating.  We note, however, that our calculations do not include the impact of 
unknown amount of depletion of electrons due to star formation as a function of redshift 
and/or mass.
With only $\la$15\% of the baryons in stars \citep{lin03} in
the local universe, we do not expect this effect to alter our results.

The level of preheating has very little effect on intermediate scales; the
SZ effect is most sensitive to gas dynamics on scales of several hundred kpc, while
non-gravitational physics is most effective on scales of tens of kpc or less. 
While non-gravitational physics can be quite effective at changing the small-scale power
in the SZ effect, it is extremely difficult to affect the overall energy budget of the
cluster in a significant way.


\section{Confusion Noise for SZ Surveys}\label{confusion}

We start by using our simulated SZ maps to investigate the SZ background as a source 
of contaminant for source counts extracted from a wide-field survey.
A glance at the maps makes it clear that the maps are highly non-Gaussian.
At the same time, the extended nature of the sources is also apparent, indicating
that the maps are also not a simple collection of points.   Consequently, the very strength of the SZ 
effect that makes it particular attractive probe of cosmology and structure formation, 
the lack of redshift-dependent dimming, also means that the SZ sky will contain, in 
addition to well-defined source signals, a background comprised of the superposition 
of numerous weak distortions associated with largely unresolved clusters spanning a 
range of redshifts and mass scales.  These background fluctuations act as a significant
contaminant for the source counts \citep{schulz03}, potentially 
distorting the observable quantities associated with identified sources and 
limiting the power of the SZ source counts as a probe of cosmology and growth
of structure since the efficacy of the measure depends on there being a well-defined 
relationship between the observable quantities (such as, flux and 
angular size) and the intrinsic properties (such as, mass) of the source population. 

As a first step towards analyzing our simulated maps, we apply a finite band-pass
filter, equivalent to an annulus in $\ell$  space, to the sky maps such that
only Fourier modes (in the flat sky approximation) with $3000<\ell<6000$ are retained.
This corresponds to an ideal CMB experiment reporting a single band power
over the specified range in $\ell$.  In principle, this approach can be 
generalized to mock up the results for specific CMB experiment by applying
the corresponding window function to the ensemble of maps. We choose a single
idealized band for the purpose of providing a concrete demonstration. 

The filtering scheme that we have adopted is a crude but
a reasonably accurate representation of the typical matched filter for upcoming
SZ surveys \citep{holder03}.  
The matched filter is designed to suppress both the large-scale
signal due to  the primary CMB fluctuations and the  very small-scale
signal largely dominated by detector noise.  With sufficiently
sensitive multifrequency surveys, the primary CMB signal could be removed 
spectrally and the large scales need not be filtered out, but the simplest 
approach would be to take advantage of the separation of spatial scales and 
use this information to remove the CMB through spatial filtering.

\begin{figure}
\begin{center}
\centerline{
\includegraphics[height=5cm]{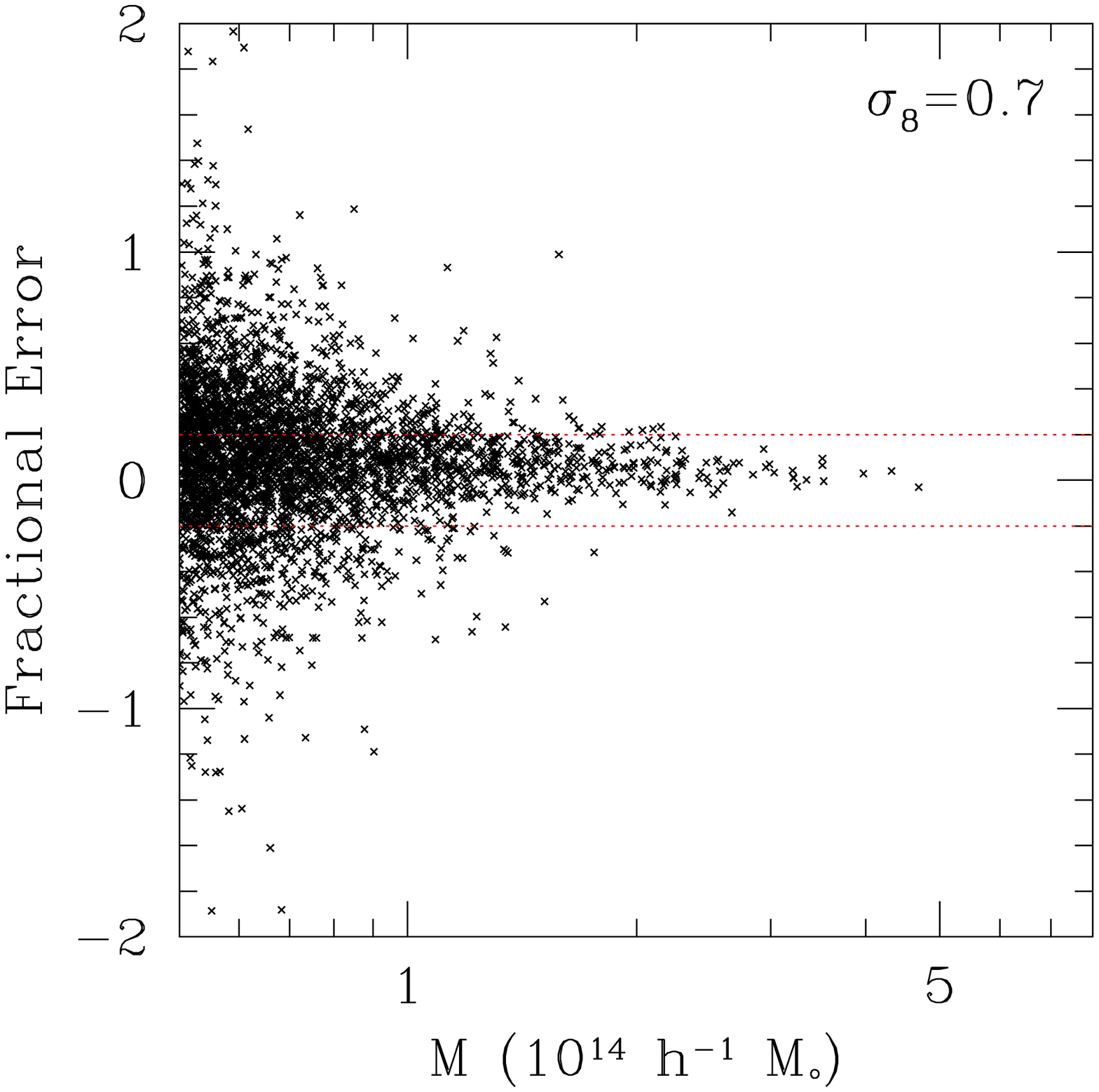}
\includegraphics[height=5cm]{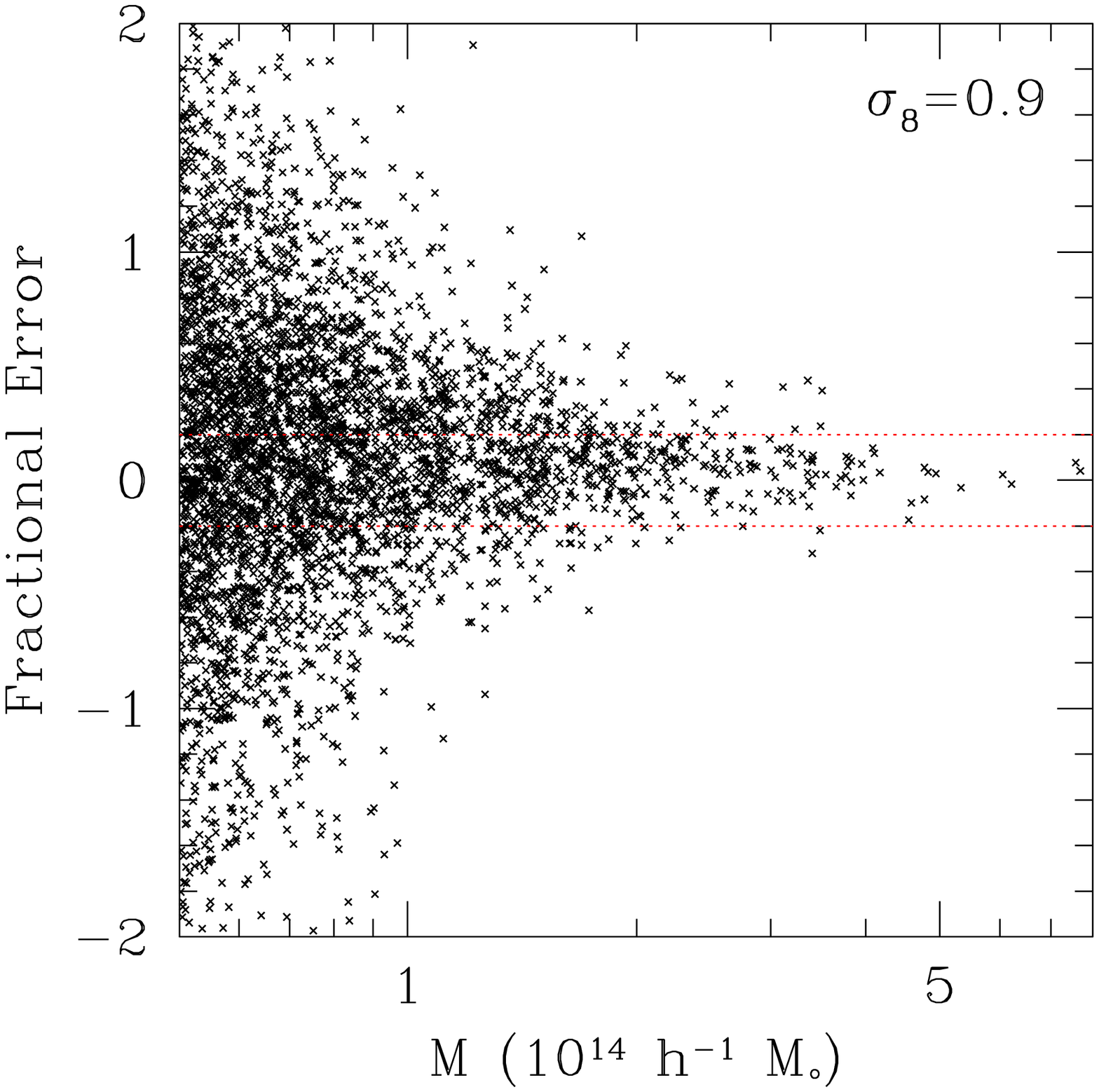}  }
\caption{
Fractional error in the SZ flux [calculated as (map-model)/model] 
in the $2^\circ x 2^\circ$ filtered sky maps at the projected position of the input halos in 
all 100  realizations of the  $S_\circ =200$,  $\sigma_8=0.9$ model (right) and 
$\sigma_8=0.7$ (left).  The dotted lines
show $\pm 20\%$ errors. Each panel shows 10000 clusters selected randomly from
the ensemble of maps. 
}
\label{fig:errors}
\end{center}
\end{figure}

In order to characterize the nature of the SZ background, we investigate the 
total SZ flux in the filtered sky maps at the projected position of each input halo
as a function of the halo mass.
In Fig. \ref{fig:errors} we show the fractional errors in the reconstructed
flux,   defined as 
\begin{equation}
\label{eqn:fracflux}
{\Delta S_\nu\over S_{\nu, true}}= {S_{\nu, map} - S_{\nu, true}\over S_{\nu, true}},
\end{equation}
where $S_{\nu, true}$ is the SZ flux expected in the filtered map
if the halo under consideration was the 
only source present, and $S_{\nu, map}$ is the measured flux in the filtered
map.  The fractional error is very
small at the projected positions of massive clusters, but the same is not true
at the other end of the mass scale.   At the projected positions of the low
mass systems, the SZ flux has significant contribution from nearby neighbours,
particularly neighbours with large footprints in the sky.   This, in turn, leads
to a very large scatter in the fractional error.  
At the risk of being repetitive, we 
stress that these maps do not include pixel noise or primary CMB fluctuation and the 
only noise here comes from sources overlapping with each other. 

At this point, the approach of this paper is a huge advantage compared to numerical
simulations. As a function of mass and redshift we can calculate the exact SZ signal
expected in the filtered maps by simply filtering the models used to generate the
maps. There is no additional noise due to substructure, triaxiality, mergers, or
large scale filamentary gas. This is a pure test of projection effects.

\begin{figure}
\begin{center}
\includegraphics[height=8cm]{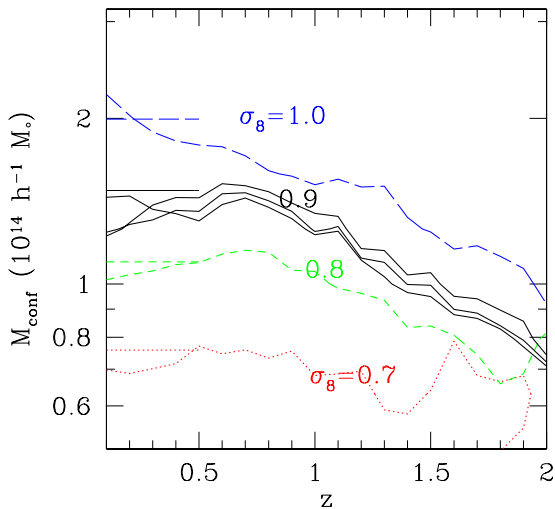}
\caption{
Mass scale below which confusion
becomes a significant contaminant (see text for details) as a function of source
redshift.   From top to bottom, the curves correspond to $\sigma_8=$ 1.0 (
dot-dashed), 0.9 (solid), 0.8 (dashed), and 0.7 (dotted).  The horizontal
lines at the left edge of the plot identifies the 
constant limiting mass that corresponds to 10 clusters per square degree in 
each model. For $\sigma_8$=0.9 there are three lines shown, corresponding
to $K_\circ=10,200,500$ keV cm$^2$.
}
\label{fig:confusion}
\end{center}
\end{figure}

We define a mass scale at a given redshift as resolved if 
the $rms$ fractional error is less than 20\%. The 
limiting mass scale below which confusion becomes important is slightly  above
$10^{14} h^{-1} M_\odot$ for the $S_\circ=200$, $\sigma_8=0.9$ model.   
As shown in
Fig. \ref{fig:confusion}, this ``confusion mass scale'' 
varies with the redshift, typically decreasing as a function of redshift,
and also scales with $\sigma_8$.  We define the confusion
mass averaged over redshift as the constant mass threshold at which the
$rms$ fractional error of SZ flux estimates above that mass is 20\%,
roughly corresponding to being comparable to the noise level for a 
$5\sigma$ detection. The confusion mass calculated this way can be well
approximated by 
\begin{equation}
M_{conf} = 10^{14}h^{-1}M_\odot (\sigma_8/0.75)^{2.5}  \ .
\end{equation}

This confusion mass scale is {\em not} very sensitive to the
details of the cluster thermodynamics.   This is because the dominant
source of confusion is due to clusters with nearly similar masses, and 
as long as the limiting mass is not close to the mass scale where 
feedback induces a qualitative change in the nature of the gas distribution
in the halos, feedback will impact halos with massess  at, just above, and 
just below the ``confusion scale'' similarly, leaving the mass scale itself
relatively unchanged.

To understand the origin of this background better, we took the simulation
volumes for our fiducial model and randomized the halo positions within 
each box.  We then repeated the mapmaking exercise using the randomized
halo positions to construct the sky maps.  The confusion levels did not,
however, change appreciably in these randomized sky maps, confirming that 
the effect described here is a true background, and not simply one caused
primarily by the correlated distribution of structure in the vicinity 
of the objects of interest.

Finally, we note that the background described above is effectively an
irreducible background, caused by structure on the same scale and 
(obviously) the same spectral behaviour as the desired objects.
For this reason, we anticipate that this background is going to be
the dominant source of noise for future generations of SZ survey
instruments.   As a rule of thumb, the confusion limit corresponds
to roughly ten objects per square degree; in this regime 
catalog construction will be straightfoward and the results robust.
This surface density is not
too dissimilar from that of typical confusion estimates in astronomy
\citep{hogg01}, which is usually quoted as of the order 25 beams per source. 
In this case, the beam is set by characterisic scale of clusters
rather than the observing beam, and is therefore only approximate.
However, 25 beams per source and a density of 10 clusters per square degree 
implies a typical ``beam'' extent (diameter) of 4' per cluster. At 
a cosmological distance this corresponds to a radial extent of 
roughly $0.5 h^{-1}$ Mpc.  The characteristic scale
of the clusters contributing to the confusion is therefore roughly half the
virial radius.

If we define the confusion mass instead
as the mass which leads to ten objects per square degree for our cosmology,
the confusion mass becomes
$M_{conf}=10^{14}h^{-1} M_\odot \times (\sigma_8/0.77)^{2.7}$, remarkably 
close to the scaling obtained from the maps. This suggests that this
really can be treated as a confusion limit and that the effective beam
size set by the galaxy cluster themselves is on the order of 4 arcminutes.

For a catalog defined by a fixed threshold relative to the detector noise, 
the scaling of $M_{conf}$ with $\sigma_8$ given above leads to bit of a 
paradox. At high values of $\sigma_8$, there are many more clusters per 
square degree and therefore, the Poisson noise is much lower.
At the same time, confusion sets in at a higher mass threshold, making the task
of constructing a catalog more challenging by increasing the rate of 
false positive detections.  At low values of $\sigma_8$, the threshold
scale for confusion is lower, allowing for the construction of very clean 
catalogs; however, because of the dearth of massive systems in low $\sigma_8$ 
cosmologies the resulting catalogs will be much more susceptible to Poisson noise.

The current generation of SZ survey experiments are expected to have noise
levels that allow imaging of clusters that are very close to the confusion
limit, on the order of $2\times 10^{14} h^{-1}M_\odot$. 
If $\sigma_8$ is near 1.0, then there is not much to be gained by improving
the sensitivity, as these experiments will then be simply imaging the SZ
background at higher signal to noise. On the other hand, if $\sigma_8$ is 
near 0.7, then the confusion mass is lower by nearly a factor of 3;
assuming that the SZ flux scales as $M^{5/3}$ (the self-similar scaling)
this translates into the confusion
limit being at a flux threshold that is nearly 5 times lower. Real experiments
have other concerns, such as radio and IR point sources, atmosphere subtraction,
detector noise, ability to do other CMB science, etc., but the cosmological
scaling of the SZ background could be an important consideration for survey
designs. 

\begin{figure}
\begin{center}
\centerline{
\includegraphics[width=50mm]{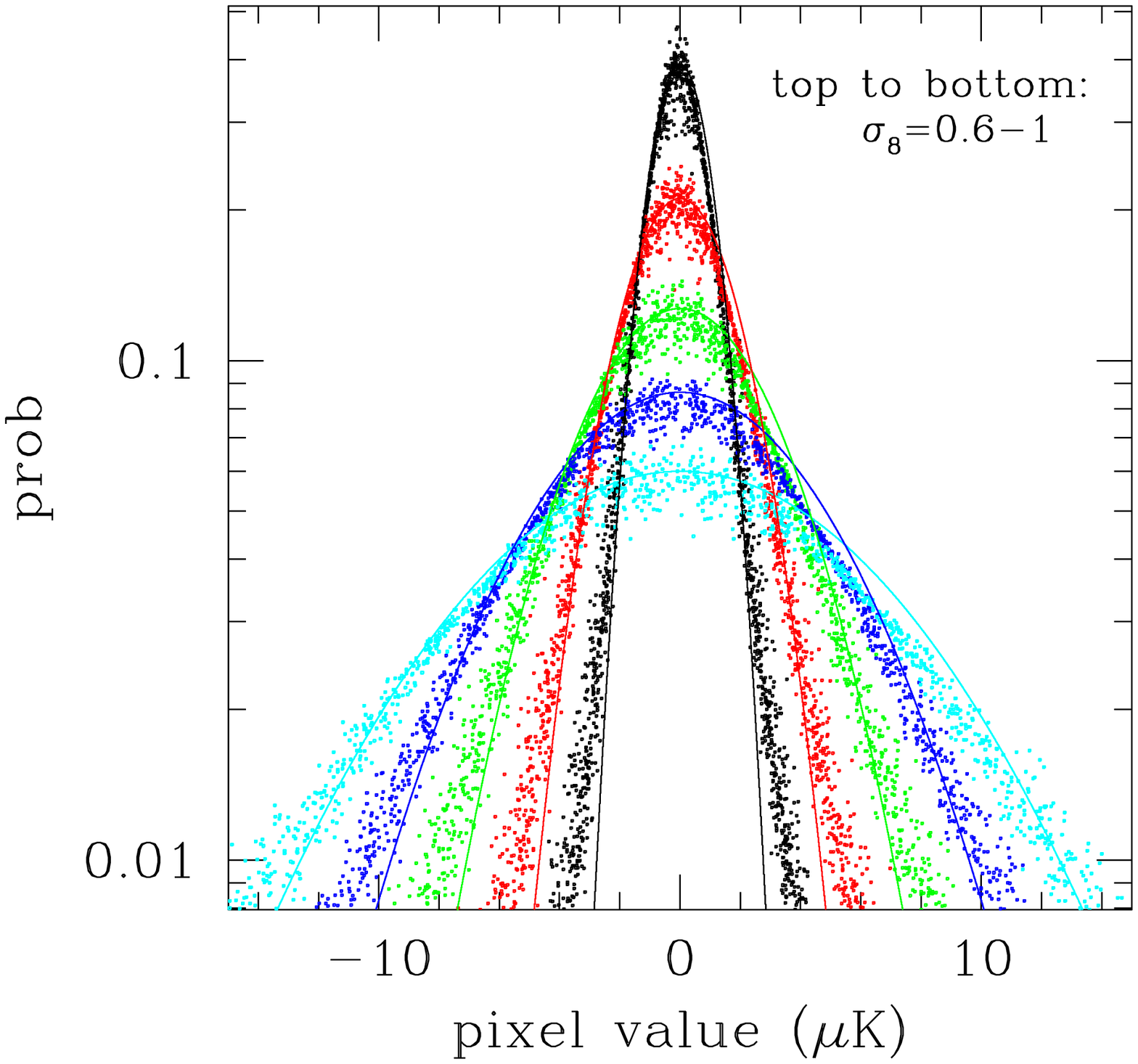}
\includegraphics[width=50mm]{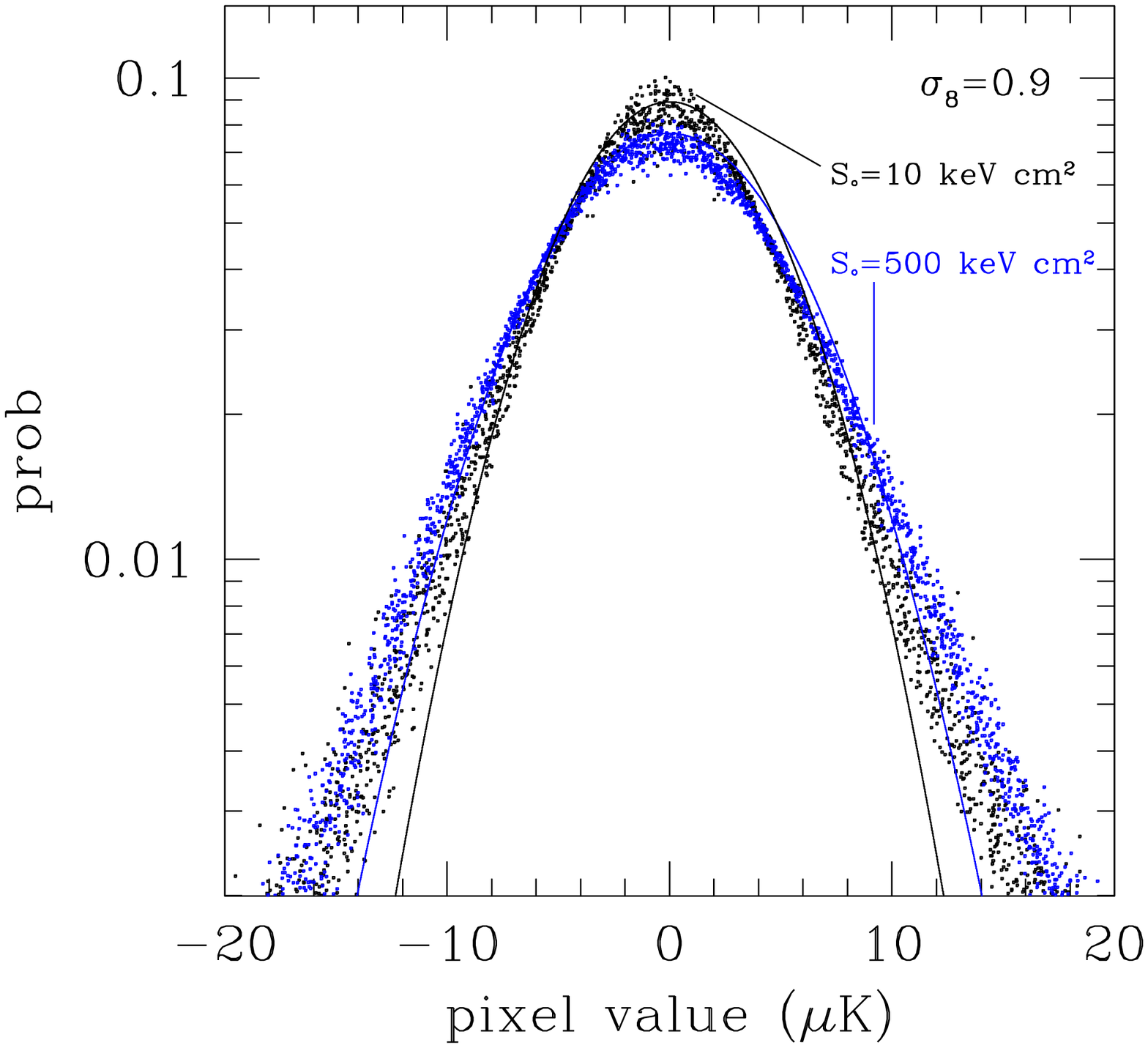} }
\caption{
Probability distribution functions of pixel values in filtered
($3000<\ell<6000$) maps for models with $S_\circ = 200$ keV cm$^2$ 
and $\sigma_8$=0.6,0.7,0.8,0.9,1.0 (top to bottom) [Left]
and for $\sigma_8=0.9$ and $S_\circ$ =10,500 keV cm$^2$ [Right].  
Results for
all 100 realizations of each model are shown.  Solid lines show 
Gaussians with the same FWHM as the mean of the 100 realizations 
for each model. }
\label{fig:pixels}
\end{center}
\end{figure}

\section{Comparison of Map Statistics}\label{mapstats}

Experiments often report the power 
in a band, which we take to be $3000<\ell<6000$ in our case.
This corresponds to the variance of a map that has been filtered in
Fourier space such that only modes within an annulus in $\ell$ space have
been retained.  The variance provides 
the complete description of the 
signal in the filtered map if the distribution of the pixel values is 
Gaussian.  In Fig. \ref{fig:pixels}, we plot the histogram of the values of 
pixels in our sky map for the each of the 100 realizations of models with
$S_\circ = 200$ keV cm$^2$ and $\sigma_8$ spans the range from 0.6 to 1.0
(top to bottom).   We will discuss this plot further in the next section.
Here, it suffices to note that the the distribution of pixel values is, in
general, not a Gaussian.
The filtered map, therefore, contains much more information than represented
by the variance, and of course, there is much more information on the sky than 
in a single filtered map.  However, we shall restrict
ourselves to analyzing sky maps filtered as described above and will
further restrict the discussion to one-point statistics.


Variance is an example of
such a measure and as noted, it completely captures all the information
in sky maps where the pixel distribution is Gaussian.   However, in the
case of non-Gaussian distributions, the variance has its limitations.
For example, the variance is very sensitive to the tails of the
distribution, much more so than, for example, the full-width 
at half-maximum (FWHM) of the distribution. 
The two measures are somewhat
complementary.  Ideally, one would like to have a simple one-point
statistical measure that is, for example, less sensitive to gas physics 
and more discrimating between cosmological models as well as an indicator
that is able to 
isolate the effects of gas physics regardless of the cosmology, etc.
Here, we investigate
the mean absolute value, FWHM, 
variance, skewness and kurtosis.  We normalize all these
statistics such that they are equivalent for Gaussian statistics:

\noindent
\begin{tabular}{ l  l }
\ \  &  \ \  \\
$\sigma_{var}=\sqrt{\langle x^2 \rangle }$ & $\hat{s}= \langle x^3 \rangle/\sigma_{var}^3$  \\
\ \  &  \ \  \\
$\sigma_{fwhm}\equiv {\rm FWHM}/2.35$  	    & $\hat{k}= \langle x^4 \rangle/\sigma_{var}^4$ \\
\ \  &  \ \  \\
$\sigma_{abs}\equiv 1.25\,\langle |x| \rangle$ & \\
\ \  &  \ \  \\
\end{tabular}

\noindent
The quantities in the left column correspond to skewness and kurtosis, respectively.
Also, we remind the reader that the filtered map 
has zero mean by construction, so there is no sensitivity to the mean 
Comptonization.


The SZ power spectrum is known to be a sensitive function of $\sigma_8$
\citep{komatsu02,komatsu99a}, so we first study the scalings of the various
one-point statistical measures with $\sigma_8$. 

As a first step, we plot in Fig. \ref{fig:pixels} the histogram of the values of 
pixels in our sky map for the each of the 100 realizations of models with
$S_\circ = 200$ keV cm$^2$ and $\sigma_8$ spanning the range from 0.6 to 1.0 in
steps of 0.1 (top to bottom).    Overall, the distribution of pixel values in the 
sky maps becomes progressively broader and more non-Gaussian as $\sigma_8$ is increased,

In the filtered maps, the effects of the entropy floor are small, as shown in the
right panel of Fig. \ref{fig:pixels}. The distributions are very similar, indicating
that the power spectrum in Fig. \ref{fig:c_ell} is giving a fair representation of
the effects of preheating.

There is a significant degree of scatter from realization to realization, 
as can be seen in Fig. \ref{fig:pixels}, due to the Poisson noise 
in the number of massive clusters.  The strongly non-Gaussian wings are also largely 
due to these massive clusters, rendering $\sigma_{var}$ an increasingly poor statistic.  
We illustrate this in Fig. \ref{fig:pixels} by plotting, for each  $\sigma_8$,
a Gaussian with the same FWHM  as the mean of the 100 realizations (solid curve).  
In addition to $\sigma_{var}$, we also consider two other one-point measures:
$\sigma_{fwhm}$ and $\sigma_{abs}$.
The behaviour of the variance and all the other measures are summarized in Table 
\ref{table:szsummary}.
All indicators scale roughly as $\sigma_i \propto \sigma_8^{n_i}$ 
with $n_i=7$, as suggested by \citep{komatsu99a}. Note that results are presented
with a proportionality symbol as a reminder that depletion of electrons through
star formation will lead to a reduction of the SZ effect. 

Interestingly, increasing the amplitude of the entropy floor drives
the distribution towards slightly less non-Gaussianity. This arises 
in part because the central Comptonization is reduced in each cluster, reducing the 
number of high-$y$ points. However,
the effect is very small. The scaling of the skewness with entropy floor
is only $\propto S_\circ^{-0.2}$ and the kurtosis only scales
as $\propto S_\circ^{-0.3}$. 

\begin{figure}
\begin{center}
\includegraphics[width=8.5cm]{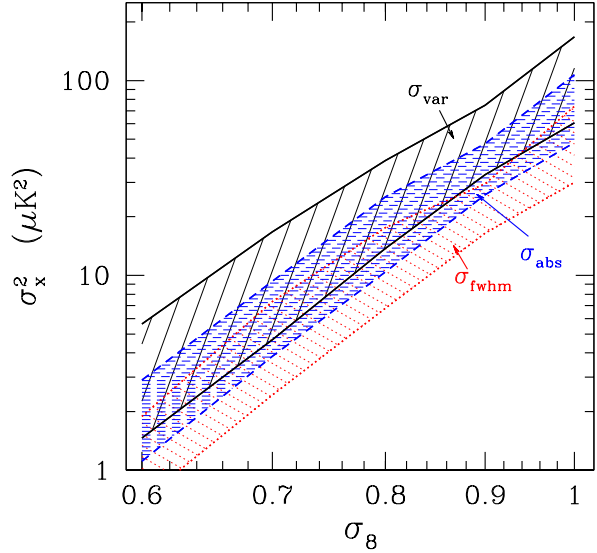}
\caption{
Distribution of variances in filtered maps
($3000 < \ell < 6000$) of size 2 degrees on a side
as a function of $\sigma_8$ assuming
fiducial gas evolution ($S_\circ=200$  keV cm$^2$. 
The 95\% intervals are shown for three
different map statistics: $\sigma_{var}^2$ (black solid), 
$\sigma_{fwhm}^2$ (red dotted), and $\sigma_{abs}^2$ (blue dashed).
}
\label{fig:var_v_sig8}
\end{center}
\end{figure}

In interpreting Table \ref{table:szsummary} it is useful to consider
the field-to-field variance of the statistical indicator, shown in the
final column for the particular case of $\sigma_8=0.9$, as well as in
Fig. \ref{fig:var_v_sig8} . As a statistical
indicator, the ideal probe would be a strong function of the parameter
of interest with small scatter. All the indicators of $\sigma$ have roughly
the same scaling with $\sigma_8$, so the scatter is the best comparison
tool. For an input $\sigma_8$ of 1, the scatter in $\sigma_{var}$
indicates that a typical map will give a 25\% estimate of this parameter,
leading to a constraint on $\sigma_8$ from a single 2x2 degree field 
that is on the order of 7\%. Using a better estimator with 15\% scatter, such as
the full-width half-maximum, would lead to an estimate on the order of 4\%. 
Of course, these estimates ignore
pixel noise and primary CMB contamination and are highly idealized. However,
using only a single ``bandpower'' the cosmological power of the SZ
background is promising.

\begin{table}
\caption{Summary of statistical indicators of filtered ($3000<\ell <6000$)
SZ maps}
\label{table:szsummary}
{
\begin{tabular}{|@{}l | r@{$\propto$}l  | r@{.}l |}
\hline
statistic 
	& \multicolumn{2}{c}{$\sigma_8$} 
	& \multicolumn{2}{c}{$\sigma_x/x$} \\
  
	& \multicolumn{2}{c}{scaling (mean)}
	& \multicolumn{2}{c}{2x2 sq deg}   \\
\hline
\hline
$\sigma_{var}^2 (\mu K^2)$ 
	& &$101\sigma_8^{7.1}$
	&~~~0&23			\\
	
$\sigma_{abs}^2 (\mu K^2)$ 
	& &$74\sigma_8^{7.2} $
	&~~~0&16				\\

$\sigma_{fwhm}^2 (\mu K^2)$ 
	& &$47\sigma_8^{7.3} $
	&~~~0&15 				\\

skewness
	&\multicolumn{2}{l}{$1.5\sigma_8^{-0.4}$}
	&~~~0&3				\\
kurtosis 
	&\multicolumn{2}{l}{$22\sigma_8^{-0.8}$}
	&~~~0&5				\\
\hline
\end{tabular}
}
\end{table}

\section{Summary and Future Work}\label{conclusions}

SZ studies face a fundamental confusion limit coming from the superposition of
many faint sources that becomes significant at a sky surface density of
roughly ten clusters per square degree. By using spherical analytic halos of known
properties we have isolated and quantified the confusion due purely to line of
sight superpositions of SZ sources.

The existence of the SZ confusion limit is a challenging limitation
to the ability to study low-mass systems with the SZ effect. Projection
effects will require that any study of systems much below $10^{14} h^{-1} M_\odot$
will be necessarily statistical in nature, through such things as
cross-correlations with optical, X-ray, or lensing maps, or stacking of
many clusters to increase the signal to noise.

The SZ confusion limit found here should not be strongly impacted
by the current uncertainty in cluster gas physics, since it is
fundamentally  determined mainly by the source density. If all clusters
were simply scaled, the relative flux between sources just above
and just below the confusion limit will not be changed. Even allowing
a mass-dependent depression or enhancement in flux will only be
important if it is more dramatic than the exponential suppression in
the mass function at high masses.

We have shown that the variance is a crude tool to
apply to SZ maps, and that better cosmological discrimination is possible
with the fwhm or the mean absolute value and that in general the one point
PDF of SZ maps is likely to be a rich dataset. There is almost certainly a
much better single number that best captures the SZ pdf as a function
of parameters, but the future will probably be better served by using a
more complete treatment of the SZ joint pdf.

The work here only looked at a single parameter, $\sigma_8$, and it would
be instructive to extend this work to include a host of other parameters,
such as $\Omega_m$ and dark energy properties, to see how different
parameters affect the SZ pdf. This is ongoing work. However, even looking
at only one parameter it is clear that the SZ background for cluster surveys
is a function of cosmology and calibrating a selection function using only a
limited set of cosmological models is dangerous.

The amount of substructure in clusters could be sensitive to
$\sigma_8$ and using ``realistic'' cluster profiles from N-body
simulations that include hydrodynamics and astrophysics will be required to understand
the details of the SZ background. At the same time, it must always
be kept in mind that the SZ background is being largely produced
by objects that are currently barely studied observationally; the properties of these
objects have not been used to tune numerical simulations.

Ultimately, the distribution of SZ values and their spatial correlations
will connect to source counts as a function of redshift, which suggests
that the SZ ``power spectrum,'' whether using the variance or some
better statistic, can benefit significantly from crude redshift estimates of all
pixels, and not just the brightest few pixels. This is under investigation; a
complete investigation of the multiwavelength characteristics of every position
on the sky is a large undertaking.

The SZ background will soon be imaged at high resolution. We have demonstrated that there
is a wealth of cosmological and astrophysical information in this signal, perhaps
most obviously evident from a glance at the maps shown in Figs. \ref{fig:skymap} 
and \ref{fig:entmap}.
This wealth of information will allow for great advances in our cosmological and
astrophysical understandings, but it is clear that the large recent advances in
instrumentation must now be matched by advanced new tools for interpreting the signals.

\vskip+2mm
This work was supported by the Natural Sciences and 
Engineering Research Council (Canada) through the Discovery Grant Awards
to GPH and AB.  GPH would also like to acknowledge support from the
Canadian Institute for Advanced Research and the the Canada Research 
Chairs Program.  IGM acknowledges support form a NSERC Postdoctoral Fellowship.
AB acknowledges support from the Leverhulme Trust 
(UK) in the form of the Leverhulme Visiting Professorship.  
Both AB and GPH would like to thank CITA and IAS for the hospitality 
shown to them during their visit during the course of this study.

\label{lastpage}

\end{document}